\documentclass[nofootinbib]{revtex4}
\usepackage{amsmath}
\usepackage{mathrsfs}
\usepackage{caption}

\begin{document}

\title{\Large BFV quantization and BRST symmetries of the gauge invariant fourth-order Pais-Uhlenbeck oscillator}

\author{Bhabani Prasad Mandal\footnote {e-mail address: bhabani.mandal@gmail.com}}

\affiliation {Department of Physics,
Banaras Hindu University, Varanasi - 221005, India.}
 
\author{Vipul Kumar Pandey\footnote {e-mail address: vipulvaranasi@gmail.com}}

\affiliation {Department of Physics and Astrophysics, 
University of Delhi, 
New Delhi - 110007, India.}

\author{Ronaldo Thibes\footnote{e-mail address: thibes@uesb.edu.br}}
\affiliation{Departamento de Ci\^encias Exatas e Naturais,
Universidade Estadual do Sudoeste da Bahia,
Itapetinga - 45700000, Brazil.}

\begin{abstract}
We perform the BFV-BRST quantization of the fourth-order Pais-Uhlenbeck oscillator (PUO).
We show that although the PUO is not naturally constrained in the sense of Dirac-Bergmann, it is possible to profit from the introduction of suitable constraints in phase space in order to obtain a proper BRST invariant quantum system.  Starting from its second-class constrained system description, we use the BFFT conversional approach to obtain first-class constraints as gauge symmetry generators.  After the Abelianization of the constraints, we obtain the conserved BRST charge, the corresponding BRST transformations and proceed further to the BFV functional quantization of the model.  We show that different possible gauge choices can be connected by finite field-dependent BRST transformations.
\end{abstract}
\maketitle
\section{Introduction} 
The Pais-Uhlenbeck oscillator (PUO) is a quantum mechanical system originally introduced in their 1950 groundbreaking paper about {\it field theories with non-localized actions} \cite{Pais:1950za}.  In their original work, Pais and Uhlenbeck were chiefly concerned with the elimination of divergent results in field theory and suggested the introduction of higher-order derivative operators. Since then, the PUO has been extensively studied and turned out to play a key central role within the scope of generalized field theories containing higher-order derivatives \cite{Lee:1969fy, Lee:1970iw, Stelle:1976gc, Stelle:1977ry, Odintsov:1991nd}.  One of the main reasons is the fact that the PUO is perhaps the simplest prototypical quantum physical system containing higher-order derivatives in which the important issues of positiveness, unitarity and causality can be isolated and studied in detail, aiming to safe applications to the more general and involved higher-derivative quantum field theories.   Particularly, in the recent past, we have seen many serious proposals within quantum gravity,  cosmology and standard model extension contexts of models based on higher-derivative theories \cite{Modesto:2015ozb, Smilga:2017arl, Ji:2019phv, Ferreira:2019lpu, Rachwal:2021bgb, Silva:2021fzh}.  The important quest of whether these models are physically healthy enough, in the sense of predictability, unitarity and causality, certainly passes through a better understanding of the corresponding details and intricacies in the simpler PUO model.

In principle, all higher-derivative models in quantum mechanics and quantum field theory were mistakenly thought to suffer from incurable instabilities leading to negative norm states corresponding to ghost propagating modes.  A first clear simple signal that that should be the case is contained in the long known Ostrogradsky instability theorem \cite{Ostrogradsky:1850fid}.  However, recent more careful analysis have definitely exorcised undesirable ghostly features and brought those models back to life.   Concerning the specific case of the fourth-order PUO, we mention the representative references \cite{Bender:2007wu, Nucci:2008ya, Nucci:2009, Mostafazadeh:2010yw, Mostafazadeh:2011qu, Pavsic:2016ykq, Raidal:2016wop}.  In \cite{Bender:2007wu}, a {\it No-Ghost Theorem for the Fourth-Order Derivative Pais-Uhlenbeck Oscillator Model} is discussed within the context of PT-symmetric quantum mechanics where it is shown to be possible to obtain unitarity and positive-norm states at the cost of giving up usual Hermicity for observables.  By preserving the point symmetries of the classical theory, Leach and Nucci have used their {\it quantizing with symmetries} road to obtain an interesting suitable alternative quantum framework for the PUO \cite{Nucci:2008ya, Nucci:2009}. 
In the couple of papers \cite{Mostafazadeh:2010yw, Mostafazadeh:2011qu}, Mostafazadeh has considered a class of transformation of variables leading to a conservative dynamical system allowing for the possibility of a complex Hamiltonian function.  The idea of imaginary-scaling counterpoints PT-symmetry and avoids the need of non-Hermitian operators.  The inclusion of interactions has been carefully considered in reference \cite{Pavsic:2016ykq}.  The relation between those different approaches has been discussed by Raidal and Veerm\"ae in a unification framework of complex higher derivative theories \cite{Raidal:2016wop}.  In the context of quantum field theory, last year, Donoghue and Menezes have clearly shown that the classical Ostrogradksy instability does not necessarily lead to a ill-defined theory at quantum level by explicit discussing a simple higher-derivative interacting model in reference \cite{Donoghue:2021eto}.

Considering this conjuncture, we may safely say that the current status of the PUO in the physics literature is indeed a positive promising one.  In this sense, we present here a novel contribution for the quantization of the PUO via the modern and well-established BRST-BFV formalism, by first constructing a consitent gauge-invariant description and discuss the effects of a finite field dependent BRST transformation.
In its original form, the PUO is not a singular constrained system in the sense of Dirac-Bergmann \cite{Bergmann:1949zz, Dirac:1950pj, Dirac}.  The presence of higher-order derivatives is not directly related to constraints in phase space as we shall explicitly show in Section II below.  In spite of that, by means of introducing auxiliary variables, Mannheim and Davidson \cite{Mannheim:2000ka, Mannheim:2004qz, Mannheim:2006rd} have obtained an equivalent description of the fourth-order PUO containing implicit second-class constraints and performed its canonical quantization as a constrained system.  The idea of the present work is then to convert those constraints from second- to first-class in order to generate gauge symmetry which allows  the construction of a correspondent BRST conserved charge at quantum level.  In this way, it is possible to obtain a consistent BRST invariant quantum theory and proceed with the BFV functional analysis constructing the quantum generating functional for all Green's functions of the theory.  Furthermore, this procedure also permits the application of the modern concept of generalized BRST transformations in which we have finite field dependent parameters \cite{sdj}.   

For the reader's convenience, this work is organized as follows.  In the next section, we introduce and briefly review the fourth-order PUO as a classical Lagrangian system.  In Section III, we show that, in spite of its higher-derivative content, the PUO is not naturally constrained in the Dirac-Bergmann sense.  By introducing extra varibles in phase space, we obtain an alternative consistent second-class Hamiltonian description of the PUO which will be used to perform the BRST-BFV quantization after the corresponding abelianization of constraints.  In Section IV, we apply the BFFT conversional method in order to obtain the first-class constraints and involutive Hamiltonian function.  In Section V, we introduce the ghost variables, obtain the quantum BRST conserved charge and write the Green's function generator in the extended phase space along the BFV functional quantization scheme.  In Section VI, we discuss the generalized finite field-dependent BRST transformations for the PUO oscillator and show how the generating functionals for different gauge choices are connected through this formalism.  We end in Section VII with some concluding remarks.
 
\section{Fourth-Order Pais-Uhlenbeck Oscillator}
Given an arbitrary field $\phi(t,\mathbf{x})$ defined in a $(1+d)$-dimensional Minkowski space governed by the fourth-order partial differential equation
\begin{eqnarray}
(\partial^2_0 - \nabla^2)(\partial^2_0 - \nabla^2 + M^2)\phi(t,\mathbf{x}) = 0
\,,
\label{emfpuo}
\end{eqnarray}
if we consider field configurations of the form $\phi = q(t) e^{i\mathbf{k}\cdot\mathbf{x}}$, then, for each particular real value of the norm of the spatial $d$-vector $\mathbf{k}$, the time-dependent function $q(t)$ satisfies
\begin{eqnarray}\label{PUode}
\frac{d^4 q}{dt^4} + (\omega^2_1 + \omega^2_2)\frac {d^2 q}{dt^2} + \omega^2_1\omega^2_2 q = 0
\label{depuo}
\end{eqnarray}
with $\omega^2_1 + \omega^2_2 = 2\mathbf{k}^2 + M^2$ and $\omega^2_1\omega^2_2 = \mathbf{k}^2(\mathbf{k}^2 + M^2)$.  
The differential equation (\ref{PUode}) characterizes the fourth-order Pais-Uhlenbeck oscillator (PUO) of frequencies $\omega_1$ and $\omega_2$ \cite{Pais:1950za}.  Note that $M = 0$ is equivalent to the equal frequency condition $\omega_1 = \omega_2$.

The usual Lagrangian function for the PUO
can be written as
\begin{eqnarray}\label{PULag}
L(q,\dot q, \ddot q) = \frac{\gamma}{2} [{\ddot q}^2 - ({\omega_1}^2+ {\omega_2}^2){\dot q^2} + {\omega_1}^2 {\omega_2}^2 q^2 ]
\label{lpuo}
\,,
\end{eqnarray}
where $\gamma$ denotes an overall multiplicative constant which can be needed for dimensional reasons.  Indeed, it can be checked that the Euler-Lagrange equation obtained directly from (\ref{PULag}) corresponds to the differential equation (\ref{depuo}).  From the classical point of view, the fourth-order ordinary differential equation (\ref{depuo}) is in fact very simple and possesses the general solutions
\begin{equation}\label{gensolA}
q(t)=A\cos\omega_1t+B\sin\omega_1t+C\cos\omega_2t+D\sin\omega_2t
\end{equation}
for the case $\omega_1\neq\omega_2$ and
\begin{equation}\label{gensolB}
q(t)=A\cos\omega t+B\sin\omega t+Ct\cos\omega t+Dt\sin\omega t
\end{equation}
for the case of equal frequencies $\omega_1=\omega_2\equiv\omega$.  In both cases the capital letters represent the rightful four arbitrary integration constants, which can be used to match the initial conditions.   The real issue concerning the PUO appears only at quantum level, namely, the important question whether it is possible to obtain a positive definite Hamiltonian operator corresponding to (\ref{PULag}) acting on a Hilbert space of positive norm states.

\section{Hamiltonian Analysis}
In order to pursue the BFV quantization of the fourth-order Pais-Uhlenbeck oscillator, in this section we obtain a suitable corresponding Hamiltonian describing a second-class constrained system.  Contrary to previous claims in the literature, the Lagrangian function (\ref{PULag}) is not naturally constrained in the sense of Dirac-Bergmann -- the presence of higher-order derivatives does not automatically grant the existence of constraints in phase space.
By definition, a singular Dirac-Bergmann system should have a null Hessian.  The Hessian of (\ref{PULag}) is given by 
\begin{equation}
W\equiv\frac{\partial^2 L}{\partial \ddot{q}^2}=\gamma
\end{equation}
which, of course, we are assuming different from zero.

As a matter of fact, the canonical Hamiltonian associated to Lagrangian (\ref{PULag}) can be immediately obtained, for instance, by using the well-known Ostrogradsky formalism for higher-order derivatives \cite{Ostrogradsky:1850fid} as follows.  Associated to $\dot q$ and $\ddot q$, introduce two momenta variables $p_1$ and $p_2$ given by
\begin{equation}
p_1=-\gamma(\omega_1^2+\omega_2^2)\dot{q}-\dot{p}_2
\end{equation}
and
\begin{equation}
p_2=\gamma\ddot{q}
\,,
\end{equation}
then, for further notation simplicity, rename the $q$ and $\dot q$ variables as
\begin{equation}
q_1\equiv q\,\mbox{~~and~~}
q_2\equiv \dot{q}
\,.
\end{equation}
Thence, the canonical Hamiltonian can be directly obtained in terms of $(q_1,q_2,p_1,p_2)$ from the Legendre transformation
\begin{equation}
H(q_i,p_i)\equiv\sum_{i=1}^{2}\dot{q}_ip_i-L(q,\dot{q},\ddot{q})\,,
\end{equation}
amounting to
\begin{equation}\label{H}
H(q_1,q_2,p_1,p_2)=q_2p_1+\frac{p_2^2}{2\gamma}
+\frac{\gamma(\omega_1^2+\omega_2^2)}{2}q_2^2
-\frac{\gamma\omega_1^2\omega_2^2}{2}q_1^2
\,.
\end{equation}
Just for a check, note that the four plain Hamilton equations associated to (\ref{H}) are given by
\begin{equation}
\begin{aligned}
&\dot{q}_1=\frac{\partial H}{\partial p_1}=q_2\,,\\
&\dot{q}_2=\frac{\partial H}{\partial p_2}=\frac{p_2}{\gamma}\,,\\
&\dot{p}_1=-\frac{\partial H}{\partial q_1}=\gamma\omega_1^2\omega_2^2q_1\,,\\
&\dot{p}_2=-\frac{\partial H}{\partial q_2}=-p_1-\gamma(\omega_1^2+\omega_2^2)q_2
\,,
\end{aligned}
\end{equation}
which are equivalent to (\ref{depuo}), as can be seen by differentiating the last one and rewriting it in terms of $q_1$ and its time derivatives by using the remaining first three ones.  Still, if the reader is not yet convinced that the Hamiltonian (\ref{H}) does not represent a constrained system in the Dirac-Bergmann sense, note that we simply performed an invertible linear transformation from the configuration space of the Lagrangian formulation to the phase space of the Hamiltonian formulation, namely
\begin{equation}
\left(
\begin{array}{c}q_1\\q_2\\p_1\\p_2\end{array}
\right)
=
\left(
\begin{array}{cccc}
1&0&0&0\\0&1&0&0\\0&-\gamma(\omega_1^2+\omega_2^2)&0&-\gamma\\0&0&\gamma&0
\end{array}
\right)
\,
\left(
\begin{array}{c}q\\\dot{q}\\\ddot{q}\\\dddot{q}\end{array}
\right)
\,.
\end{equation}
and we can easily come back to configuration space whenever we wish.
Furthermore, as an additional bonus, the Ostrogradsky Hamiltonian function (\ref{H}) has the nice property of being a conserved quantity representing the total energy of the system, albeit being neither positive definite nor bounded.

Although the system (\ref{PULag}) is not constrained and possesses the well-defined canonical Ostrogradsky classical Hamiltonian (\ref{H}), it can be alternatively described by an equivalent Dirac-Bergmann constrained system, as has been recently shown in references \cite{Mannheim:2004qz, Mannheim:2006rd, Bender:2007wu}.   At first, the introduction of constraints at this point may seem rather artificial and clumsy.  However, the idea here concerns reducing the order of the time derivatives and being able to use the whole modern machinery of quantization of constrained systems to shed light on the open problem of obtaining a correct quantization of the Pais-Uhlenbeck oscillator.  In particular, this approach can be used to produce a consistent gauge-invariant version description for the PUO.  Therefore, building on the previous work of Mannheim, Davidson and Bender \cite{Mannheim:2004qz, Mannheim:2006rd, Bender:2007wu}, in the next paragraphs we show how the Pais-Uhlenbeck oscillator can be described as a second-class constrained system in order to, in the following remaining sections, proceed with the conversion of the constraints to first-class and perform its corresponding BRST-BFV quantization. 

Aiming to reduce the order of the time derivatives present in (\ref{PULag}), while still
granting the independence of $q$ and $\dot q$ as dynamical coordinates, we consider a new coordinate $x$ and impose through the equations of motion that it should actually equal $\dot q$.  This can be done by means of a Lagrange multiplier variable $\lambda$ directly introduced in the Lagrangian.  More precisely, we consider the alternative Lagrangian function
\begin{eqnarray}
L_c(q,x,\lambda,\dot{q},\dot{x},\dot{\lambda}) = \frac{\gamma}{2} [{\dot x}^2 - ({\omega_1}^2+ {\omega_2}^2){x^2} + {\omega_1}^2 {\omega_2}^2 q^2 ] + \lambda (\dot q - x )
\,,
\label{lpuolm}
\end{eqnarray}
which exhibits lower-order time derivatives as (\ref{PULag}) at the cost of possessing more variables.   In the recent literature, the use of the reduction of order technique to study higher-derivative models from different perspectives has been shown to often produce new important insights \cite{Thibes:2016ivt, Nogueira:2018jdm, Dai:2020qpc}.
The Lagrangian (\ref{lpuolm}) is equivalent to the higher-order previous one (\ref{PULag}) as can be checked for instance by the equations of motion and, lo and behold, does characterize a constrained system in the Dirac-Bergmann sense -- its corresponding Hessian matrix is clearly singular.  To construct the phase space, we introduce canonically
conjugated momenta $p_x$, $p$ and $p_\lambda$, corresponding respectively to the three coordinate variables $x$, $q$ and $\lambda$, defined as
\begin{eqnarray}
p_x \equiv \frac {\partial L_c }{\partial \dot x} = \gamma \dot x, \quad p \equiv \frac {\partial L_c }{\partial \dot q} = \lambda, \quad p_{\lambda} \equiv \frac {\partial L_c }{\partial \dot \lambda} = 0 
\,.
\label{conm}
\end{eqnarray}
Using relations (\ref{conm}) and performing a Legendre transformation in (\ref{lpuolm}), a canonical Hamiltonian can be written as
\begin{eqnarray}
H_c &=& p_x \dot x + p\dot q + p_\lambda \dot \lambda - L_c\,, \nonumber\\
&=& \frac{p^2_x}{2\gamma} + \frac{\gamma}{2}(\omega^2_1 + \omega^2_2) x^2 - \frac{\gamma}{2} \omega^2_1 \omega^2_2 q^2 + \lambda x
\,.
\label{hfpuo}
\end{eqnarray}

For the next steps, we apply the well-known Dirac-Bergmann algorithm for constrained systems \cite{Dirac, Sundermeyer:1982gv, Gitman:1990qh, Henneaux:1992ig}.  Associated to the above Hamiltonian (\ref{hfpuo}), we have two primary constraints, namely,
\begin{equation}
\Omega_1 \equiv {p} - \lambda\,,
~~~\mbox{ and }~~~ 
\Omega_2 \equiv p_{\lambda}
\,,
\label{prmcon}
\end{equation} 
which can be linearly added to (\ref{hfpuo}) multiplying two yet undetermined Lagrange multiplier functions, respectively $u_1$ and $u_2$, to produce the Primary Hamiltonian
\begin{eqnarray}
H_P &=& \frac{p^2_x}{2\gamma} + \frac{\gamma}{2}(\omega^2_1 + \omega^2_2) x^2 - \frac{\gamma}{2} \omega^2_1 \omega^2_2 q^2 + \lambda x + u_1(p - \lambda) + u_2 p_\lambda
\,.
\label{htpuo}
\end{eqnarray}
Defining the usual Poisson brackets in terms of the phase space variables $(q,x,\lambda,p,p_x,p_\lambda)$, the time evolution of the constraints (\ref{prmcon}) in terms of the Primary Hamiltonian (\ref{htpuo}) is given by equations
\begin{equation}\label{evprcn1}
\{\Omega_1, H_P\} = \gamma \omega^2_1 \omega^2_2 q - u_2 + \Omega_1\{\Omega_1, u_1\} + \Omega_2\{\Omega_1, u_2\} %\approx 0
\,,
\end{equation}
and
\begin{equation}
\{\Omega_2, H_P\} = - x + u_1 + \Omega_1\{\Omega_2, u_1\} + \Omega_2\{\Omega_2, u_2\} %\approx 0
\,.
\label{evprcn2}
\end{equation}
Imposing the stability of relations (\ref{prmcon}) within the primary constraints hypersurface, the Lagrange multiplier functions can be determined as
\begin{eqnarray}
u_1 = x\,\mbox{~~and~~}  u_2 = \gamma \omega^2_1 \omega^2_2 q 
\,,
\end{eqnarray}
which substituted back into equation (\ref{htpuo}) give us the final expression for total Hamiltonian
\begin{equation}\label{HT}
H_T = \frac{p^2_x}{2\gamma} + \frac{\gamma}{2}(\omega^2_1 + \omega^2_2) x^2 - \frac{\gamma}{2} \omega^2_1 \omega^2_2 q^2 + p x + \gamma \omega^2_1 \omega^2_2 q p_\lambda
\,.
\end{equation}
No further constraints are generated by the Dirac-Bergmann consistency algorithm and we note that we have a genuine second-class constrained system. 
For operatorial quantization purposes, we note that the complete set of second class constraints $\Omega_a$, $a=1,2$, generates a Dirac bracket Lie algebra structure in phase space.  From the Dirac bracket general definition
\begin{equation}\label{DB}
\left\lbrace F, G \right\rbrace^* \equiv \{ F, G \}
-\{ F,\Omega_a \} \{\Omega_a,\Omega_b\}^{-1} \{\Omega_b, G\}
\,,
\end{equation} for arbitrary phase space functions $F$ and $G$, we find the fundamental
non-null Dirac brackets among the fundamental variables
\begin{equation}
\{q,\lambda\}^*=\{q,p\}^*=\{x,p_x\}^*
=1
\,.
\end{equation}
In the next section, we convert the second-class constraints to first-class in order to apply the BFV functional quantization scheme.

\section{Construction of First-Class Constraints and Hamiltonian}
To convert the constraints from second to first class, we shall use the standard Batalin-Fradkin-Fradkina-Tyutin (BFFT) conversional approach \cite{Batalin:1986aq, Batalin:1986fm, Egorian:1988ss, Batalin:1989dm, Batalin:1991jm, Pandey:2021myh}.    As a first step, we introduce a pair of auxiliary fields $\varphi^a$, $a = 1,2$, satisfying the extended Poisson bracket relations
\begin{equation}
\omega^{ab}\equiv \{\varphi^a, \varphi^b\} = \epsilon^{ab}
\end{equation}
where $\epsilon^{ab}$ denotes the totally antisymmetric Levi-Civita symbol with the convention $\epsilon^{12}=1$.
Following references \cite{Batalin:1991jm, Pandey:2021myh}, in order to calculate the first order correction to the constraints (\ref{prmcon}), we look for a solution of $X_{ab}$ in  
\begin{eqnarray}\label{XwX}
\Delta_{ab} +  X_{ac}\omega^{cd}X_{bd} = 0
\,,
\end{eqnarray}
where
\begin{equation}
\Delta_{ab}\equiv\{\Omega_a,\Omega_b\}
\,
\end{equation}
denotes the constraints matrix.
Now, using relations (\ref{prmcon}), in terms of $a,b = 1,2$, equation (\ref{XwX}) explicitly means
\begin{eqnarray}
&&a = 1\,,\, b = 1 \longrightarrow X_{11}X_{12} - X_{12}X_{11} = 0\,,\nonumber\\
&&a = 1\,,\, b = 2 \longrightarrow X_{11}X_{22} - X_{12}X_{21} = 1\,, \nonumber\\
&&a = 2\,,\, b = 1 \longrightarrow X_{21}X_{12} - X_{22}X_{11} = - 1 \,,\nonumber\\
&&a = 2\,,\, b = 2 \longrightarrow X_{21}X_{22} - X_{22}X_{21} = 0\,,
\end{eqnarray}
from which we pick the simple possible solution
\begin{equation}
\begin{array}{c}
X_{11} = 0\,, ~~
X_{12} = - 1 = - X_{21}  ~~\mbox{ and }~~
X_{22} = 0\,.
\end{array}
\end{equation}
This choice leads to the constraints first order correction
\begin{equation}\label{Phi_a}
\begin{array}{l}
\Omega_1~~\longrightarrow~~{\Phi}_1 = \Omega_1 + \Omega_1^{(1)} = p -\lambda -\varphi^2\,,
\\
\Omega_2~~\longrightarrow~~{\Phi}_2 = \Omega_2 + \Omega_2^{(1)} = p_\lambda + \varphi^1\,,
\end{array}
\end{equation}
and it can already be verified that 
\begin{eqnarray}
\{{\Phi}_1,  {\Phi}_2\} = 0\,.
\end{eqnarray}
Therefore, for the present case, the constraints first order correction is sufficient to achieve an Abelian algebra.\footnote{Since the initial constraints are linear in the phase space variables, this result was already expected.  See for instance reference \cite{Amorim:1995sh}.}

Next, for obtaining a gauge invariant Hamiltonian for the system, we consider the general expression for the modified Hamiltonian \cite{Batalin:1991jm}
\begin{equation}\label{calH}
{\cal H}=\sum_n H^{(n)}
\,,
\end{equation}
with
\begin{equation}
H^0\equiv H_T
\,,
\end{equation}
\begin{equation}
H^{(n+1)} \equiv - \frac{1}{n+1}\varphi^a\omega_{ab}X^{bc}(q,p)G_c^{(n)}
\label{nocHm}
\,,
\end{equation}
and 
\begin{equation}\label{Ga}
G_a^{(n)} \equiv \sum_{m = 0}^{n}\{\Omega_a^{(n-m)}, H^{(m)} \}_{(q,p)} + \sum_{m = 0}^{n-2}\{\Omega_a^{(n-m)}, H^{(m+2)}\}_{(\varphi)} + \{\Omega_a^{(n+1)}, H^{(1)}\}_{(\varphi)}
\,.
\end{equation}  Note that $H^{(n)}$ represents the order $n$ in the fields $\varphi^a$ correction to $H$.
Inserting the inverse of the matrices $\omega^{ab}$ and $X_{ab}$, given 
respectively by 
\begin{equation}
\omega_{ab} = 
\left(
\begin{array}{cc}
~0 & -1 \\
~1 & ~0 \\
\end{array}
\right)
~~~~
\mbox{ and }
~~~~
X^{ab} = 
\left(
\begin{array}{cc}
~0 & ~1 \\
-1 & ~0 \\
\end{array}
\right)
\,,
\end{equation}
into equation (\ref{nocHm}), we obtain the first order corrections to the Hamiltonian (\ref{HT}) due to fields $\varphi^1$ and $\varphi^2$ respectively
as
\begin{equation}
H^{(1)}_{\varphi^1} = -\varphi^1\omega_{11}X^{11}G_1^{(0)} -\varphi^1\omega_{11}X^{12}G_2^{(0)} -\varphi^1\omega_{12}X^{21}G_1^{(0)} -\varphi^1\omega_{12}X^{22}G_2^{(0)}
= \varphi^1\gamma\omega^2_1 \omega^2_2 p_\lambda
\label{FodHcr}
\,,
\end{equation}
and
\begin{equation}
H^{(1)}_{\varphi^2} = -\varphi^2\omega_{21}X^{11}G_1^{(0)} - \varphi^2\omega_{21}X^{12}G_2^{(0)} -\varphi^2\omega_{22}X^{21}G_1^{(0)} -\varphi^2\omega_{22}X^{22}G_2^{(0)} = 0 
\,,
\end{equation}
where, from (\ref{Ga}), we have used $G_1^{(0)} = \{\Omega_1, H_T \}_{(q,p)} = \gamma\omega^2_1 \omega^2_2 p_\lambda$ and $G_2^{(0)} = \{\Omega_1, H_T \}_{(q,p)}   = 0$. 
Similarly the second order corrections to Hamiltonian are given by
\begin{equation}
H^{(2)}_{\varphi^1} = -\varphi^1\omega_{11}X^{11}G_1^{(1)} -\varphi^1\omega_{11}X^{12}G_2^{(1)} -\varphi^1\omega_{12}X^{21}G_1^{(1)} -\varphi^1\omega_{12}X^{22}G_2^{(1)}
= \frac{1}{2}{\varphi^1}^2\gamma\omega^2_1 \omega^2_2 
\label{sodHcr}
\end{equation}
and
\begin{equation}
H^{(2)}_{\varphi^2} = -\varphi^2\omega_{21}X^{11}G_1^{(1)} - \varphi^2\omega_{21}X^{12}G_2^{(1)} -\varphi^2\omega_{22}X^{21}G_1^{(1)} -\varphi^2\omega_{22}X^{22}G_2^{(1)} = 0 
\,,
\end{equation}
where $G_1^{(1)} = \{\Omega_1, H^{(1)} \}_{(q,p)} = \varphi^1\gamma\omega^2_1 \omega^2_2$ and $G_2^{(1)} = \{\Omega_2, H^{(1)} \}_{(q,p)} = 0$.
There are no higher order correction due to the fields $\varphi^1$ and $\varphi^2$ since all higher order values for $G_{a}^n$, with $n\geq2$, identically vanish.
So, the final form for the modified Hamiltonian (\ref{calH}) can be written as
\begin{equation}
{\cal H} = \frac{p^2_x}{2\gamma} + \frac{\gamma}{2}(\omega^2_1 + \omega^2_2) x^2 - \frac{\gamma}{2} \omega^2_1 \omega^2_2 q^2 + p x + \gamma \omega^2_1 \omega^2_2 q p_\lambda + \varphi^1\gamma\omega^2_1 \omega^2_2 p_\lambda + \frac{1}{2}{\varphi^1}^2\gamma\omega^2_1 \omega^2_2 
\label{htpuof}
\,.
\end{equation}
Now, we can easily verify that the converted constraints are involutive with the modified total Hamiltonian of the system
\begin{equation}
\{{\cal H},  {\Phi}_a\} = 0
\,,
\end{equation}
and we have achieved this section's goal of obtaining a first-class system through its constraints Abelianization governed by a fully involutive Hamiltonian.

\section{BFV-BRST quantization of the Pais-Uhlenbeck Oscillator}
Resuming from the obtained first-class system,
in this section, we perform the BRST-BFV \cite{Becchi:1974xu, Becchi:1975nq, Tyutin:1975qk, Batalin:1977pb, Fradkin:1975cq} quantization of the fourth-order Pais-Uhlenbeck oscillator.   For notation convenience, we rename the two BFFT variables $\varphi^a$ respectively to $\varphi$ and $\pi$ 
and rewrite the first-class Hamiltonian (\ref{htpuof}) as
\begin{equation}\label{calH}
{\cal H}=
\frac{p_x^2}{2\gamma}
+
\frac{\gamma(\omega_1^2+\omega_2^2)x^2}{2}
-
\frac{\gamma}{2}\omega_1^2\omega_2^2(q+\varphi)^2
+xp+\gamma\omega_1^2\omega_2^2(q+\varphi)(p_\lambda+\varphi)
\,.
\end{equation}
Since this represents a genuine first-class theory, we introduce two additional Lagrange multipliers $v^a$, along with their respective conjugated momenta $w_a$, corresponding to the first class constraints (\ref{Phi_a}).  Following the usual BRST-BFV quantization formalism, in order to explicitly realize the BRST symmetry, we extend further the phase space by defining a set of  odd Grassmannian parity
ghost fields  $({\cal C}^a, \bar{\cal C}_a)$ and respective momenta $(\bar{\cal P}_a ,{\cal P}^a)$, associated to the first class constraints $\Phi_a$, satisfying the canonical relations 
\begin{equation}
\lbrace {\cal P}^a, \bar{\cal C}_b \rbrace
=
\lbrace {\cal C}^a, \bar{\cal P}_b \rbrace
=-\delta^a_b
\,.
\end{equation}
Besides, 
we also introduce a conserved ghost-number operator with eigenvalues according
\captionsetup{labelformat=empty}
\begin{table}[ht]
\centering
{\small\begin{tabular}{lrcccc} \hline\hline\rule{0pt}{3ex} &$\,\,\,z$&$\,\,\,{\cal C}^a$&$\,\,\bar{\cal C}_a$&$\,\,{\cal P}^a$&$\,\,\bar{\cal P}_a$\\
\hline
Grassmann parity&$0$&$\phantom{+}1$&$\phantom{+}1$&$\phantom{+}1$&$\phantom{+}1$\\
Ghost number&$0$&$+1$&$-1$&$+1$&$-1$
\\
\hline\hline
\end{tabular}}
\caption{Table I - Grassmann parity and ghost numbers} \label{table1}
\end{table}
to Table \ref{table1}
in which $z$ collectively denotes all even Grassmannian phase space variables, i.e.,
\begin{equation}
z=(q,p,x,p_x,\lambda,p_\lambda,v^a,w_a)
\,.
\end{equation}

At this point, we are ready to define the nillpotent BRST charge
\begin{equation}\label{Omega}
{\cal Q}={\cal C}^a\Phi_a -i {\cal P}^a w_a\,,
\end{equation}
as the generator of the BRST symmetry.  More precisely, if $F$ denotes an arbitrary function defined in the extended phase space, its BRST transformation generated by (\ref{Omega}) is given by
\begin{equation}\label{sF}
sF = \lbrace F, {\cal Q} \rbrace
\,.
\end{equation}
Thus, from (\ref{sF}), the extended phase variables non-null BRST variations read explicitly
\begin{equation}\label{BRS}
\begin{aligned}
&sq=s p_\lambda =-s\varphi={\cal C}^1\,,\\
&s\lambda=-s\pi={\cal C}^2\,,\\
&sv^a=-i{\cal P}^a\,,\\
&s{\bar{\cal C}}_a = i w_a\,,\\
&s{\bar{\cal P}}_1 = -p + \lambda + \pi\,,\\
&s{\bar{\cal P}}_2 = -p_\lambda -\varphi\,.
\end{aligned}
\end{equation}
As can be easily checked, the BRST transformation (\ref{BRS}) is nillpotent and represents a symmetry of the Hamiltonian (\ref{calH}).

Now the BFV quantization of the model can be obtained in terms of a given gauge-fixing fermionic function $\Psi$  by defining the generating functional as
\begin{equation}\label{Z}
Z_\Psi=\int [d\sigma] \exp \big\lbrace -\frac{i}{\hbar} S_{ext} \big\rbrace
\end{equation}
where $[d\sigma]$ denotes the functional integration measure 
\begin{equation}
[d\sigma]={\cal D}q\,{\cal D} p\, {\cal D} x\, {\cal D} p_x\, {\cal D} \lambda  \, {\cal D } p_\lambda\, {\cal D} v^a\,{\cal D} w_a
{\cal D C}^a\,{\cal D}\bar{\cal C}_a \, {\cal D{\cal P}}^a\, {\cal D}\bar{\cal P}_a
\end{equation}
and $S_{ext}$ stands for the extended action given by 
\begin{equation}\label{Sext}
S_{ext}=\int_{t_i}^{t_f}dt\,
\left(
\dot{q}p+\dot{x}p_x+\dot{\lambda}p_\lambda +\dot{\varphi}\pi + \dot{v}^aw_a+\dot{\cal P}^a \bar{\cal C}_a + \dot{\cal C}^a \bar{\cal P}_a 
-{\cal H} +\lbrace \Psi,{\cal Q} \rbrace
\right)
\,.
\end{equation}
Note that the extended action is of first order in the time derivative and depends on the specific gauge through the gauge-fixing fermion $\Psi$ present in (\ref{Sext}).  In spite of that, the effective action can also be shown to be invariant under the BRST transformations (\ref{BRS}).  Indeed, applying the BRST operator $s$ to $S_{ext}$, we readily obtain
\begin{equation}
s\,S_{ext}=
\int_{t_i}^{t_f}dt\,
\frac{d}{dt}
\left(
\lambda {\cal C}^1 - \varphi {\cal C}^2
\right) = 0
\,.
\end{equation}
Furthermore, the Fradkin-Vilkovisky theorem \cite{Fradkin:1975cq, Batalin:1977pb,  Henneaux:1985kr}
 assures us that the generating functional (\ref{Z}) is in fact gauge-independent.  A useful standard form for $\Psi$ is given by
\begin{equation}\label{PSI}
\Psi = i\bar{\cal C}_a\chi^a+\bar{\cal P}_a w^a
\end{equation}
where $\chi^a$ denote two gauge functions independent of the ghost variables.  For this form, using the BRST charge (\ref{Omega}), we have
\begin{equation}\label{psiom}
\lbrace \Psi,{\cal Q} \rbrace
=
-w_a\chi^a + i\bar{\cal C}_a\lbrace \chi^a, \Phi_b \rbrace {\cal C}^b
+\bar{\cal C}_a\lbrace \chi^a, w_b \rbrace {\cal P}^b
-v^a \Phi_a -i\bar{\cal P}_a{{\cal P}}^a
\,.
\end{equation}
A possible interesting natural gauge choice, which leads to an effective quantum action without time derivatives for the ghost fields is given by
\begin{equation}\label{GC1}
\begin{cases}
\chi^1=\dot{v}^1+\frac{1}{2}\omega_1\omega_2\varphi\,,  \\
\chi^2=\dot{v}^2+\frac{1}{2}\omega_1\omega_2\pi\,.
\end{cases}
\end{equation}
In fact, substituting (\ref{GC1}) into (\ref{psiom})
and integrating the generating functional (\ref{Z}) in $\bar{\cal P}_a$, ${{\cal P}}^a$,  $w_a$, $\varphi$ and $\pi$, we get the intermediate action
\begin{eqnarray}\label{Sext'}
S'_{ext}&=&\int_{t_i}^{t_f}dt\,
\bigg(
\dot{q}p+\dot{x}p_x+\dot{\lambda}p_\lambda
-v^1(p-\lambda)-v^2p_\lambda
-i\omega_1\omega_2{\bar{\cal C}}_a{\cal C}^a
\nonumber\\&&
 -\frac{p_x^2}{2\gamma}-\gamma\frac{(\omega_1^2+\omega_2^2)x^2}{2}
 +\frac{\gamma}{2}\omega_1^2\omega_2^2q^2-xp-\gamma\omega_1^2\omega_2^2qp_\lambda
\bigg)
\,.
\end{eqnarray}
Then,
integrating further in $v^a$, $p_\lambda$ and $\lambda$ we obtain the final quantum effective action
\begin{eqnarray}\label{Seff}
S_{eff}&=&\int_{t_i}^{t_f}dt\,
\Big(
\dot{q}p+\dot{x}p_x
-i\omega_1\omega_2{\bar{\cal C}}_a{\cal C}^a
\nonumber\\&&
 -\frac{p_x^2}{2\gamma}-\gamma\frac{(\omega_1^2+\omega_2^2)x^2}{2}
 +\frac{\gamma}{2}\omega_1^2\omega_2^2q^2-xp
\Big)
\,.
\end{eqnarray}
On the other hand, by using  the following  gauge choice
\begin{equation}
\begin{cases}
\chi^1=\omega_1\omega_2\varphi \,,  \\
\chi^2=\omega_1\omega_2\pi \,,
\end{cases}
\end{equation}
and performing a similar calculation, it is possible to obtain an alternative quantum effective action with a dynamical term for the ghost fields given by
\begin{eqnarray}
{S'}_{eff}&=&\int_{t_i}^{t_f}dt\,
\left(
\dot{q}p+\dot{x}p_x+\dot{\lambda}p_\lambda +\dot{\varphi}\pi + \dot{v}^aw_a+i{\cal \dot{\bar{ C}}}_a{\dot{\cal C}}^a
\right.\nonumber\\&&\left.
-i\omega_1\omega_2{\cal C}_a{\cal C}^a
-{\cal H} -\omega_1\omega_2w_1\varphi-\omega_1\omega_2w_2\pi-v^a\Phi_a
\right)\label{S'}
\,.
\end{eqnarray}
Comparing to usual field theory of continuous variables, the second quantum effective action (\ref{S'}) would correspond to a covariant gauge allowing for the possibility of combining the ghost time derivatives with eventual space derivatives. In the next section we will show how different generating functionals for the different gauge choices are connected through finite field dependent transformations.

\section{Finite Field BRST Transformation for the Pais-Uhlenbeck Oscillator}
In this section, we generalize the nilpotent BRST symmetry constructed in the previous one following the work of Joglekar and Mandal \cite{sdj}. In that seminal work, the usual BRST transformation, which is characterized by a infinitesimal, anticommuting and global parameter, is generalized to have the transformation parameter finite and field dependent without affecting the symmetry of the effective action. This type of generalized BRST transformations are known as finite field dependent BRST  (FFBRST) transformations. However, a FFBRST transformation does not leave the path integral measure invariant precisely due to finiteness of the transformation parameter.
Under a certain condition, the non-trivial Jacobian caused by the FFBRST transformation of the path integral measure is expressed as a local functional of the fields, which eventually modifies the effective action of the theory \cite{sdj}. Due to this remarkable feature, the FFBRST transformation is capable of relating the generating functionals corresponding to different effective actions. The FFBRST formulation has found various applications in gauge field theories over the years \cite{ffbrst,ff1,ff2,ffbrst3,ffbrst4,ffbrst5,ffbrst6,ffbrst7,ffbrst8,ffbrst9,ffbrst10,ffbrst11,ffbrst12,ffbrst13,ffbrst14,ffbrst15,ffbrst16,ffbrst17,ffbrst18,ffbrst19,ffbrst20}. As an application of FFBRST transformation, we would like to show how  the generating functionals corresponding to two different effective actions of quantized Pais-Uhlenbeck oscillators in two different gauges are also connected through a FFBRST transformation.

For this purpose we briefly recapitulate the techniques of FFBRST formulation. As the first step, all the fields (generically denoted as $\sigma $) are made to depend on a numerical parameter $k$ ($0\leq k \leq 1$),  in such a fashion that $\sigma(k=0) = \sigma $ is the initial field and $\sigma(k = 1) = \sigma' $ is the transformed one. Considering the infinitesimal parameter field dependent the usual BRST transformation  is written as
 \begin{equation}\label{inb}
 d\sigma(k)=s\sigma(k)\ \Theta^\prime(\sigma(k),k)dk
 \,.
 \end{equation}
 Here $ \Theta^\prime(\sigma(k),k)dk $ is an infinitesimal but field dependent Grassmann parameter.  A FFBRST transformation $\sigma \rightarrow\sigma^\prime $ is then constructed by integrating equation (\ref{inb}) from $k=0$ to $k=1$ as
  \begin{equation}\label{ffb}
  \sigma^\prime  = \sigma + s\sigma \Theta(\sigma)
 \end{equation}
 where 
  \begin{equation}
  \Theta(\sigma ) = \int_0^1\Theta^\prime(\sigma(k),k)dk
  \,.
 \end{equation}
The FFBRST transformation in equation (\ref{ffb}) leaves the Faddeev-Popov effective action invariant but  the path integral measure changes non-trivially under such finite transformation. The non-trivial Jacobian which is the source of all new results is written as
 \begin{equation}\label{jac}
 [d\sigma(k)] = J(k) [d\sigma^\prime(k)]
 \end{equation}
where $J(k=0)=1$. It has been shown in reference \cite{sdj} that this Jacobian $J(k)$ can be replaced by a local functional of the fields within the functional integral as
\begin{equation}\label{jk}
J(k)= e^{i S_1(\sigma(k),k)}
 \end{equation}
if and only if
 \begin{eqnarray}
\int [d\sigma(k)]\big[\frac{1}{J(k)}\frac{d J(k)}{d k}-i\frac{dS_1}{dk}\big] e^{i (S_1 + S_{eff})} = 0 
\label{ffbc}
\end{eqnarray}
holds. Here, $\frac{dS_1}{dk}$ is a total derivative of $S_1$ with respect to $k$ in which the dependence on $\sigma(k)$ is also differentiated. The Jacobian change is then calculated as 
 \begin{eqnarray}
\frac{J(k)}{J(k+dk)} &=& \Sigma_{\sigma}{\pm}\frac{\delta \sigma(k+dk)}{\delta \sigma(k)}\nonumber\\
                     &=& 1 - \frac{1}{J(k)}\frac{d J(k)}{d k} d k
\label{jacbc}
\end{eqnarray}
where the symbol ${\pm}$ denotes a positive/negative sign for bosonic/fermionic fields ($\sigma$), respectively.
Prior to functional integration, the effective action for the PUO using the BFV formulation can be written using equations (\ref{calH}), (\ref{Sext}) and (\ref{psiom}) as
\begin{eqnarray}
S_{eff} &=&\int_{t_1}^{t_2}  dt \big[{\dot q} p + {\dot x}p_x + {\dot\varphi} p_\varphi + {\dot\lambda}p_\lambda + {\dot v^a}w_a + {\dot{\cal C}^a} {\bar {\cal P}}_a + \dot{{\cal P}^a} {\bar {\cal C}}_a - \frac{p^2_x}{2\gamma} - \frac{\gamma}{2}(\omega^2_1 + \omega^2_2)x^2 \nonumber\\ &+& \frac{\gamma}{2}\omega^2_1\omega^2_2 (q + \varphi)^2 - xp - \gamma\omega^2_1\omega^2_2 (q + \varphi)(p_\lambda + \varphi) -w_a\chi^a + i\bar{\cal C}_a\lbrace \chi^a, \Phi_b \rbrace {\cal C}^b
+\bar{\cal C}_a\lbrace \chi^a, w_b \rbrace {\cal P}^b
\nonumber\\ &-&v^a \Phi_a -i\bar{\cal P}_a{{\cal P}}^a\big] 
\label{qfet} 
\end{eqnarray}
being consistently BRST invariant under the transformation (\ref{BRS}). The finite version of the BRST transformation in equations (\ref{BRS}) can then be written as
\begin{equation}\label{ffbt}
\begin{aligned}
&\delta q =\delta p_\lambda =-\delta\varphi={\cal C}^1\Theta (\sigma)\,,\\
&\delta \lambda =-\delta\pi={\cal C}^2\Theta(\sigma)\,,\\
&\delta v^a =-i{\cal P}^a\Theta(\sigma)\,,\\
&\delta{\bar{\cal C}}_a = i w_a\Theta(\sigma)\,,\\
&\delta {\bar{\cal P}}_1 = (-p + \lambda + \pi)\Theta(\sigma)\,,\\
&\delta {\bar{\cal P}}_2 = (-p_\lambda -\varphi)\Theta(\sigma)\,,
\end{aligned}
\end{equation}
with all
other variables having null BRST variations and $\Theta$ being a finite field dependent, global and anti-commuting parameter. It is straightforward to check that the effective action given in equation (\ref{qfet}) is invariant under this FFBRST  transformation. Now we consider the possibility of different gauge choices.
\subsection{I set of Gauge choices} 
We make  the first gauge choice as
\begin{equation}
\chi^1 = \dot {v^1} + \omega_1 \omega_2\varphi, \quad \chi^2 = \dot {v^2} + \omega_1 \omega_2\pi \,.
\label{gchc1} 
\end{equation}
The  effective action under this gauge condition is then written as
\begin{eqnarray}
S_{eff}^I &=&\int_{t_1}^{t_2}  dt \big[{\dot q} p + {\dot x}p_x + {\dot\varphi} p_\varphi + {\dot\lambda}p_\lambda + {\dot v^a}w_a + {\dot{\cal C}^a} {\bar {\cal P}}_a + \dot{{\cal P}^a} {\bar {\cal C}}_a - \frac{p^2_x}{2\gamma} - \frac{\gamma}{2}(\omega^2_1 + \omega^2_2)x^2 \nonumber\\ &+& \frac{\gamma}{2}\omega^2_1\omega^2_2 (q + \varphi)^2 - xp - \gamma\omega^2_1\omega^2_2 (q + \varphi)(p_\lambda + \varphi) - w_1(\dot {v^1} + \omega_1 \omega_2\varphi) - w_2 (\dot {v^2} + \omega_1 \omega_2\pi)  \nonumber\\ &+& i\bar{\cal C}_a(-2\omega_1\omega_2) {\cal C}^a
+\bar{\cal C}_1\lbrace \dot{v^1}, w_1 \rbrace {\cal P}^1 + \bar{\cal C}_2\lbrace \dot{v^2}, w_2 \rbrace {\cal P}^2
- v^1 (p - \lambda - \pi) - v^2 (p_\lambda + \varphi) -i\bar{\cal P}_a{{\cal P}}^a\big] 
\,.
\label{actgc1} 
\end{eqnarray}
 
For another set of gauge choice
\begin{eqnarray}
\chi^1 = \omega_1 \omega_2\varphi, \quad \chi^2 = \omega_1 \omega_2\pi\,,
\label{gchc2} 
\end{eqnarray}
the  effective action is written as
\begin{eqnarray}
S_{eff} ^{II}&=&\int_{t_1}^{t_2}  dt \big[{\dot q} p + {\dot x}p_x + {\dot\varphi} p_\varphi + {\dot\lambda}p_\lambda + {\dot v^a}w_a + {\dot{\cal C}^a} {\bar {\cal P}}_a + \dot{{\cal P}^a} {\bar {\cal C}}_a - \frac{p^2_x}{2\gamma} - \frac{\gamma}{2}(\omega^2_1 + \omega^2_2)x^2 \nonumber\\ &+& \frac{\gamma}{2}\omega^2_1\omega^2_2 (q + \varphi)^2 - xp - \gamma\omega^2_1\omega^2_2 (q + \varphi)(p_\lambda + \varphi) - w_1( \omega_1 \omega_2\varphi) - w_2 (\omega_1 \omega_2\pi)  \nonumber\\ &+& i\bar{\cal C}_a(-2\omega_1\omega_2) {\cal C}^a
+\bar{\cal C}_1\lbrace \dot{v^1}, w_1 \rbrace {\cal P}^1 + \bar{\cal C}_2\lbrace \dot{v^2}, w_2 \rbrace {\cal P}^2
- v^1 (p - \lambda - \pi) - v^2 (p_\lambda + \varphi) -i\bar{\cal P}_a{{\cal P}}^a\big] 
\label{actgc2} 
\end{eqnarray}
 
Now, we know that the gauge fixing and ghost parts of the Lagrangian always are written as BRST exact term as follows
\begin{equation}
({\cal L}_{gf} + {\cal L}_{gh}) = - i s (\bar{\cal C}_a \chi^a)
\,.
\label{gfgh} 
\end{equation}
The right hand side further can be written
for the gauge choice I as
\begin{equation}
= - i s (\bar {\cal C}_1 (\dot {v^1} + \omega_1 \omega_2\varphi) + \bar{\cal C}_2 (\dot {v^2} + \omega_1 \omega_2\pi))
\label{gfgh1} 
\end{equation}
and for gauge choice II as
\begin{equation}
= - i s (\bar {\cal C}_1 ( \omega_1 \omega_2\varphi) + \bar{\cal C}_2 (\omega_1 \omega_2\pi))
\label{gfgh2} 
\,.
\end{equation}
Now we will construct an appropriate  FFBRST  transformation to establish  the connection between the generating functionals corresponding to these two effective actions explicitly.
For that, we choose the finite BRST parameter $\Theta^\prime$  as
\begin{eqnarray}
\Theta^\prime= i \gamma'\int dt [{\bar{\cal C}}_a {\dot v}^a ]
\label{thtp1} 
\,.
\end{eqnarray} 

Here $\gamma'$ is an arbitrary constant and all the fields depend on $k$.
The infinitesimal change in the Jacobian corresponding to this choice of the FFBRST parameter is calculated using equation (\ref{jacbc}) as
\begin{equation}
 \frac{1}{J(k)}\frac{d J(k)}{d k} = - i \gamma'\int dt [- iw_a{\dot v}^a - i {\dot{\cal P}}^a {\bar {\cal C}}_a]
 \label{cjcb}
 \,.
\end{equation}
Now we will make an ansatz for the local functional of fields $S_1$ by considering all possible terms that could arise from such a transformation as
\begin{eqnarray}
S_1 &=& \int dt [i\xi_1 w_1 {\dot v}^1  + i\xi_2 w_1 {\omega}_1{\omega}_2\varphi + i\xi_3 w_2 {\dot v}^2 + i\xi_4 w_2 {\omega}_1{\omega}_2\pi + i\xi_5{\dot{\cal P}}^1 {\bar{\cal C}}_1 + \xi_6 \omega_1\omega_2 {\cal C}^1 {\bar {\cal C}}_1 \nonumber\\ &+& i \xi_7 {\dot{\cal P}}^2 {\bar {\cal C}}_2 + \xi_8 \omega_1\omega_2 {\cal C}^2 {\bar{\cal C}}_2]
\label{ansso}
\,,
\end{eqnarray}
where $\xi_i(k)$ are $k$ dependent arbitrary parameter with the initial condition $\xi_n (k = 0) = 0$.  
To satisfy the condition in equation (\ref{ffbc}), we calculate
\begin{eqnarray}
\frac{dS_1}{dk} &=& \int dt [i\xi'_1 w_1 {\dot v}^1  + i\xi'_2 w_1 {\omega}_1{\omega}_2\varphi + i\xi'_3 w_2 {\dot v}^2 + i\xi'_4 w_2 {\omega}_1{\omega}_2\pi + i\xi'_5{\dot {\cal P}}^1 {\bar{\cal C}}_1 + \xi'_6 \omega_1\omega_2 {\cal C}^1 {\bar{\cal C}}_1 \nonumber\\ &+& i \xi'_7 {\dot{\cal P}}^2 {\bar {\cal C}}_2 + \xi'_8 \omega_1\omega_2 {\cal C}^2 {\bar{\cal C}}_2 + \Theta'\{ i\xi_1 w_1 ({-i\dot{\cal P}}^1)  + i\xi_2 w_1 {\omega}_1{\omega}_2(-{\cal C}^1) + i\xi_3 w_2 (-i{\dot {\cal P}}^2) + i\xi_4 w_2 {\omega}_1{\omega}_2(-{\cal C}^2)  \nonumber\\ &+& i\xi_5{\dot{\cal P}}^1 ({i w}_1) + \xi_6 \omega_1\omega_2 ({-iw}^1){\cal C}^1 + i \xi_7 {\dot {\cal P}}^2 ({iw}_2) + \xi_8 \omega_1\omega_2 ({-iw}_2){\cal C}^2\}]
\label{csowk}
\end{eqnarray}
where $\xi'_n = \frac {d\xi_n}{dk}$.
Now we substitute the results of equations (\ref{csowk}) and (\ref{cjcb}) into condition (\ref{ffbc}) to obtain
\begin{eqnarray}
 &\displaystyle\int  [d\sigma] &  \exp[i (S_I[\sigma(k)] + S_1[\sigma(k), k] )]\int dt [( -\gamma' + \xi'_1)w_1 {\dot v}^1  + \xi'_2  w_1 {\omega}_1{\omega}_2\varphi + (-\gamma' + \xi'_3) w_2 {\dot v}^2 +\xi'_4 w_2 {\omega}_1{\omega}_2\pi \nonumber\\ &&+ (-\gamma'+ \xi'_5){\dot{\cal P}}^1 {\bar{\cal C}}_1 +\xi'_6\omega_1\omega_2 {\cal C}^1 {\bar{\cal C}}_1 + (-\gamma'+ \xi'_7){\dot{\cal P}}^2 {\bar{\cal C}}_2 + \xi'_8 \omega_1\omega_2 {\cal C}^2 {\bar {\cal C}}_2 + \Theta' \{(-\xi_1 + \xi_5){\dot {\cal P}}^1 ({i w}_1)  \nonumber\\ &&+(\xi_2 + \xi_6)\omega_1\omega_2 ({-iw}^1){\cal C}^1 + (-\xi_3 + \xi_7){\dot {\cal P}}^2 ({iw}_2) + (\xi_4 + \xi_8)\omega_1\omega_2 ({-iw}_2){\cal C}^2 \} ] = 0
 \label{ffbrc}
 \,.
\end{eqnarray} 

The terms proportional to $\Theta'$, which are nonlocal due to $\Theta'$, vanish independently  if
\begin{equation}
-\xi_1 + \xi_5 = 0, \quad \xi_2 + \xi_6 = 0, \quad -\xi_3 + \xi_7 = 0, \quad \xi_4 + \xi_8 = 0 \,.
\label{reiji}
\end{equation}
To make the remaining local terms in (\ref{ffbrc}) vanish, we need the following conditions:
\begin{eqnarray}
-\gamma + \xi'_1 &=&0, \quad -\gamma + \xi'_3 =0, \quad -\gamma + \xi'_5 = 0, \quad -\gamma + \xi'_7 = 0\,,\nonumber\\
\xi'_2 &=&0, \quad \xi'_4 = 0, \quad \xi'_6 = 0, \quad \xi'_8 = 0\,.
\label{rbjg}
\end{eqnarray}
The differential equations for $\xi_n(k)$ can be solved with the initial conditions $\xi_n(0) = 0$ to obtain the solutions
\begin{equation}
\xi_1 = \gamma' k, \quad  \xi_3 = \gamma' k,\quad \xi_5 = \gamma' k,\quad \xi_7 = \gamma' k,\quad \xi_2 = \xi_4 = \xi_6 = \xi_8 = 0 \,.
\label{voj}
\end{equation}

Putting the values of these parameters into the expression for $S_1$ and choosing the arbitrary parameter $\gamma' = 1$, we obtain
\begin{eqnarray}
S_1[\sigma(k=1), k=1] = i\int dt [w_1 {\dot v}^1 + w_2 {\dot v}^2 + {\dot{\cal P}}^1 {\bar{\cal C}}_1 + {\dot{\cal P}}^2 {\bar{\cal C}}_2 ] 
\label{yeso}
\,.
\end{eqnarray}
Now we can see that 
\begin{equation}
S_{eff}^I + S_1 =  S_{eff}^{II}
\end{equation}

Thus the FFBRST transformation with the finite parameter $\Theta$ that is defined by (\ref{thtp1}) changes the generating functional $Z_{Gauge I} $  to $Z_{Gauge II} $\begin{eqnarray}
Z_{Gauge I} &=& \int [d\sigma (x,k)] \exp (iS^I_{eff}[\sigma (x,k)] )\stackrel{FFBRST}{\longrightarrow}
\int [d\sigma']\exp[i{ (S^I_{eff}[\sigma'] + S_1[\sigma'] )}] \nonumber \\
&=&\int [d\sigma]\exp[i {(S^I_{eff}[\sigma] + S_1[\sigma , 1] )}] 
= \int [d\sigma ] \exp (iS^{II}_{eff}[\sigma] ) \equiv Z_{Gauge II}
\label{gflc}
\,.
\end{eqnarray}
This connection between generating functions can be established for any two gauges through FFBRST transformation. Next section we consider another set of choices.
\subsection{II set of gauge choices}
We make following two choices of gauge

Gauge choice III:
\begin{eqnarray}
\chi^1 = \dot {v^1} + \omega_1 \omega_2\varphi, \quad \chi^2 = \omega_1 \omega_2\pi
\label{gchc3} 
\,.
\end{eqnarray}

Gauge choice IV:
\begin{eqnarray}
\chi^1 = \omega_1 \omega_2\varphi, \quad \chi^2 = {\dot v}^2 + \omega_1 \omega_2\pi
\label{gchc4} 
\,.
\end{eqnarray}

The actions under these two gauges are written respectively as
\begin{eqnarray}
S_{eff}^{III} &=&\int_{t_1}^{t_2}  dt \big[{\dot q} p + {\dot x}p_x + {\dot\varphi} p_\varphi + {\dot\lambda}p_\lambda + {\dot v^a}w_a + {\dot{\cal C}^a} {\bar{\cal P}}_a + \dot{{\cal P}^a} {\bar{\cal C}}_a - \frac{p^2_x}{2\gamma} - \frac{\gamma}{2}(\omega^2_1 + \omega^2_2)x^2 \nonumber\\ &+& \frac{\gamma}{2}\omega^2_1\omega^2_2 (q + \varphi)^2 - xp - \gamma\omega^2_1\omega^2_2 (q + \varphi)(p_\lambda + \varphi) - w_1(\dot {v^1} + \omega_1 \omega_2\varphi) - w_2 (\omega_1 \omega_2\pi)  \nonumber\\ &+& i\bar{\cal C}_a(-2\omega_1\omega_2) {\cal C}^a
+\bar{\cal C}_1\lbrace \dot{v^1}, w_1 \rbrace {\cal P}^1 - v^1 (p - \lambda - \pi) - v^2 (p_\lambda + \varphi) -i\bar{\cal P}_a{{\cal P}}^a\big] 
\label{actgc3} 
\end{eqnarray}
and 
\begin{eqnarray}
S_{eff}^{IV} &=&\int_{t_1}^{t_2}  dt \big[{\dot q} p + {\dot x}p_x + {\dot\varphi} p_\varphi + {\dot\lambda}p_\lambda + {\dot v^a}w_a + {\dot{\cal C}^a} {\bar{\cal P}}_a + \dot{{\cal P}^a} {\bar{\cal C}}_a - \frac{p^2_x}{2\gamma} - \frac{\gamma}{2}(\omega^2_1 + \omega^2_2)x^2 \nonumber\\ &+& \frac{\gamma}{2}\omega^2_1\omega^2_2 (q + \varphi)^2 - xp - \gamma\omega^2_1\omega^2_2 (q + \varphi)(p_\lambda + \varphi) - w_1( \omega_1 \omega_2\varphi) - w_2 ({\dot v}^2 +\omega_1 \omega_2\pi)  \nonumber\\ &+& i\bar{\cal C}_a(-2\omega_1\omega_2) {\cal C}^a
+ \bar{\cal C}_2\lbrace \dot{v^2}, w_2 \rbrace {\cal P}^2
- v^1 (p - \lambda - \pi) - v^2 (p_\lambda + \varphi) -i\bar{\cal P}_a{{\cal P}}^a\big] 
\label{actgc4} 
\,.
\end{eqnarray}
 
The gauge fixing and ghost parts of the action can be written for the gauge choice III as
\begin{equation}
= - i s (\bar {\cal C}_1 (\dot {v^1} + \omega_1 \omega_2\varphi) + \bar {\cal C}_2 (\omega_1 \omega_2\pi))
\label{gfgh3} 
\end{equation}
and for the gauge choice IV as
\begin{equation}
= - i s(\bar {\cal C}_1 ( \omega_1 \omega_2\varphi) + \bar {\cal C}_2 (\dot {v^2} + \omega_1 \omega_2\pi))
\label{gfgh4} 
\,.
\end{equation}

In this case, we need to construct a FFBRST transformation with finite BRST parameter  $\Theta^\prime$  as
\begin{eqnarray}
\Theta^\prime= i \gamma'\int dt [{\bar{\cal C}}_1 {\dot v}^1 - {\bar {\cal C}}_2 {\dot v}^2 ]
\label{thtp2}
\,. 
\end{eqnarray} 

Here $\gamma'$ is an arbitrary constant and all the fields depend on $k$.
Following exactly the same procedure as in the earlier case, we obtain the
Jacobian factor as $e^{i\tilde{S}_1}$, with $\tilde{S}_1$  given by
\begin{eqnarray}
S_1[\sigma(k=1), k=1] = i\int dt [w_1 {\dot v}^1 - w_2 {\dot v}^2 + {\dot{\cal P}}^1 {\bar {\cal C}}_1 - {\dot{\cal P}}^2 {\bar{\cal C}}_2 ] 
\label{yeso}
\,,
\end{eqnarray}
where we have chosen arbitrary parameter $\gamma' = 1$.
Note again that, due to our appropriate construction of FFBRST transformation, we get
\begin{equation}
S_{eff}^{III} + S_1 =  S_{eff}^{IV}
\end{equation}
and the FFBRST transformation with the finite parameter $\Theta$ defined by equation (\ref{thtp2}) changes the generating functional $Z_{III}$ as
\begin{eqnarray}
Z_{ Gauge III} &=& \int [d\sigma] \exp (iS_{eff}^{III}[\sigma] )\stackrel{FFBRST}{\longrightarrow}
\int [d\sigma']\exp[i{ (S^{III}_{eff}[\sigma'] + S_1[\sigma', 1] )}] \nonumber\\
&=&\int [d\sigma]\exp[i {(S^{III}_{eff}[\sigma] + S_1[\sigma ] )}] 
= \int [d\sigma ] \exp (iS^{IV}_{eff}[\sigma] ) \equiv Z_{ Gauge IV}
\label{gflc}
\,.
\end{eqnarray}

\section{Conclusion}
We have discussed the functional BFV quantization of the PUO and its corresponding BRST and FFBRST symmetries.  As we have seen in the Introduction, after a deeper analysis, higher-derivative models in quantum mechanics and quantum field theory have proven not to be necessarily ill-defined and have experienced a revival in the recent physics literature.  In this sense, we have been able to obtain here for the first time the BRST symmetries of the fourth-order PUO.  We have shown that it is possible to describe the PUO as a Dirac-Bergmann second-class constrained system and use this fact as a bridge to construct a BRST conserved charge in the extended phase space.  The reduction of order of the PUO was achieved by means of the introduction of auxiliary variables and that led to natural second-class constraints in phase space which were treated through the Dirac-Bergmann consistency algorithm.  This framework allowed us to apply the BFFT constraints conversion approach, turning the constraints from second- to first-class.  That was precisely the key to obtain the standard BRST symmetries of the model and proceed with its BFV quantization.  We have further obtained the generating functional of the theory, with an extended quantum action including the corresponding ghost fields.  Finally, we have shown how to connect different gauges by means of the FFBRST transformations.  The new FFBRST transformations generalize the usual BRST ones by means of employing a finite field dependent parameter and have been applied to different gauge models along the last years, their successful application to the PUO confirms its strength as an important tool in the general functional quantization framework.

{\bf Acknowledgements:} One of us (BPM) acknowledges the Research Grant for Faculty under IoE Scheme (number 6031).

\end{document}